\documentstyle[12pt,aaspp4]{article}

\def\etal{{\it et al.}}
\def\eg{{\it e.g.,}}

\begin{document}

\title{A Companion Galaxy to the Post-Starburst Quasar UN\,J1025$-$0040}

\author {Gabriela Canalizo, Alan Stockton,}

\affil{Institute for Astronomy, University of Hawaii, 2680 Woodlawn
 Drive, Honolulu, HI 96822}

\author {M. S. Brotherton, Wil van Breugel}
\affil{Institute of Geophysics and Planetary Physics,
Lawrence Livermore National Laboratory, 7000 East Avenue, P.O. Box 808, L413,
Livermore, CA 94550}

\begin{abstract}
UN\,J1025$-$0040 is a quasar at $z = 0.6344$ that shows an extremely bright 
post starburst population of age $\sim 400$ Myr 
(Brotherton \etal\ 1999\markcite{bro99}).   
Images of UN\,J1025$-$0040 show a nearly stellar object 4\farcs2 SSW of the 
quasar.  We present imaging and spectroscopy that confirm that this object 
is a companion galaxy at redshift $z = 0.6341$.   We estimate an age of
$\sim 800$ Myr for the dominant stellar population in the companion.
The companion appears to be interacting with the quasar host galaxy, and this
interaction may have triggered both the starburst and the quasar activity
in UN\,J1025$-$0040.

\end{abstract}


\section{Introduction}

Although the relationship between starbursts and nuclear activity remains
controversial (see, \eg\ Joseph 1999\markcite{jos99} and Sanders 
1999\markcite{san99}), the two are often
found together.  Furthermore, there has long been strong circumstantial 
evidence that QSO activity is often triggered by interactions or mergers
(Stockton 1999\markcite{sto99} and references therein)
and it is well established that virtually all of the most luminous 
starbursts show strong interaction (Sanders \& Mirabel 1996\markcite {san96});
at least some of these also show nuclear activity.  
Indeed, some objects classified as ultraluminous infrared
galaxies (ULIGs) on the basis of their far-IR flux densities were originally
known as QSOs (\eg\ 3C\,48, Mrk\,231, Mrk\,1014).   Spectroscopy of hosts
and companions of these and other objects that show both QSO and ULIG 
characteristics confirm that they almost universally have dominant
post-starburst populations (eg, Boroson \& Oke 1984\markcite{bor84},
 Stockton, Canalizo \& Close 1998\markcite{sto98},  Canalizo \& Stockton 
2000a\markcite{can00a}, 2000b\markcite{can00b}).   

One of the most spectacular examples of such an object is
UN\,J1025$-$0040, recently identified by Brotherton \etal\ 
(1999\markcite{bro99}; hereafter Paper I), originally 
targeted as a quasar candidate by the 2dF 
survey\footnotemark\footnotetext{http://msowww.anu.edu.au/$\sim$rsmith/QSO\_Survey/qso\_surv.html} 
(Smith et al. 1996\markcite{smi96}). UN\,J1025$-$0040 is  
a quasar where, at least in the optical, the flux contribution from
a recent massive starburst is closely balanced with that from the AGN, 
leading to an unusual composite spectrum.  
Brotherton \etal\ estimate an age of 400 Myr for the
post-starburst population.  They emphasize the possibility that
UN\,J1025$-$0040 may be a transitional object between ULIGs and the classical
QSO population. 

In the deep $K_S$ image shown as Fig.~2 in Paper I\markcite{bro99}, there is a faint object
4\farcs2 south-southwest of the quasar, having an essentially stellar profile. 
However, the host galaxy of the quasar is elongated in roughly the same 
direction, so, as suggested in Paper I\markcite{bro99}, this object might be a companion 
galaxy, rather than simply an intervening star.
Here we present spectroscopic and imaging observations that confirm
this suggestion.

\section{Observations and Data Reduction}

Spectroscopic observations of UN\,J1025$-$0040 and its companion were carried
out on UT 1999 April 22 with the Low Resolution Imaging Spectrometer (LRIS;
Oke \etal\ 1995\markcite{oke95}) on the Keck II telescope.  The slit was 
1\arcsec\ wide
and oriented at a PA of 23\arcdeg\ to pass through both the quasar and
the companion; the grating had 400 grooves mm$^{-1}$ and was blazed at
8500 \AA , giving a dispersion of 1.86 \AA\ pixel$^{-1}$, and a projected
slit width of $\sim8.5$ \AA. The observations were taken close to the
meridian, so the slit position angle was always within 30\arcdeg\ of the
parallactic angle, and the zenith angle was never more than 21\arcdeg.
There was no order-separating 
filter in the beam, so we use only the portions of the spectrum
uncontaminated by second order overlap.
The total integration time was 2160 s.

The spectroscopic reduction followed standard  procedures. After
subtracting bias and dividing by a normalized halogen lamp flat-field frame,
we rectified the individual frames and placed them on a wavelength scale
by a transformation to match measurements of the spectrum of a Hg-Ne-Kr
lamp. We then traced the spectra of the quasar and the companion using
routines in the IRAF {\it apextract} package.
We calibrated the spectra by observations of the spectrophotometric
standard stars Feige 34 and Wolf 1346 (Massey \etal\ 1988\markcite{mas88}), 
and averaged the one-dimensional traces
from the three frames with the IRAF task {\it scombine}.

We obtained
$K_{S}$ images of UN\,J1025$-$0040 using the Near-Infrared
Camera (NIRC; Matthews and Soifer 1994\markcite{mat94}) at the Keck I 
telescope on UT 1998 April 18.  The details of the 
observations and data reduction are described in Paper I\markcite{bro99}.

We obtained $H$-band images of UN\,J1025$-$0040 with the University of Hawaii
(UH) 2.2 m telescope on UT 1999 April 3. We used the $1024\times1024$ QUIRC 
(HgCdTe) infrared camera (Hodapp \etal\ 1996\markcite{hod96}) at f/31, which 
gives an image
scale of 0\farcs0608 pixel$^{-1}$. The sky conditions were photometric and 
the seeing was $\sim$0\farcs3. We used four UKIRT faint standard stars 
(Casali \& Hawarden 1992\markcite{cas92}) for flux calibration.  The total 
integration time was 3600 s.

We also obtained $R$ and $I$-band images with the UH 2.2 m telescope on 
UT 1999 April 6--7 and UT 1999 May 5.  The April images were obtained through
thin cirrus with $\sim$0\farcs9 seeing and were calibrated to the May images,
for which conditions were photometric with $\sim$0\farcs7 seeing. All images 
were obtained with a Tektronix $2048\times2048$
CCD with an image scale of 0\farcs22 pixel$^{-1}$ and were calibrated by
5 standard stars in Selected Area 101 (Landolt 1992\markcite{lan92}).  The 
total exposures on the UN\,J1025$-$0040 field were 4500 s in $R$ and 9000 s 
in $I$.

\section{Results}

Figure 1 shows the spectra of UN\,J1025$-$0040 and the companion object,
which clearly is a galaxy associated with the quasar.   The redshift of 
the starburst in the host galaxy, as measured from stellar absorption lines, 
is $z_{host} = 0.6344 \pm 0.0001$.  The redshift of the companion galaxy 
$z_{comp} = 0.6341 \pm 0.0001$ was measured from the [\ion{O}{2}] 
$\lambda$3727 emission line.

We compared the spectrum of the companion to Bruzual \& Charlot 
(1996\markcite{bru96}) isochrone synthesis models.   The lighter line in 
Fig.~1 is an 800 Myr old instantaneous burst, Scalo (1986\markcite{sca86})
initial mass function, 
solar metallicity model, which gives a reasonable fit to the data.   
Note that, in spite of the rather noisy spectrum, the absorption features
in the model are fit well by the observed spectrum.  The agreement of the
absorption features and overall shape of the continuum in the model with the
observed spectrum, as well as the presence of the [\ion{Ne}{3}] emission line 
at $\lambda$3869, provide further evidence that the galaxy is at redshift  
$z = 0.6341$. 

The images of UN\,J1025$-$0040 in $R$, $I$, and $H$ bands (Fig.~2) show the 
same basic morphology as the $K_{s}$ image shown in Fig.~2 of 
Paper I\markcite{bro99}.   In general,
the images of the host galaxy show an elongation towards the companion,
and possibly a hint of a bridge between the two objects.   
Our 2-D spectra (Fig.~3) show faint and clumpy [\ion{O}{2}] $\lambda$3727 
emission between the companion and the host.

Table~1 gives the photometry of UN\,J1025$-$0040 and its companion.
Figure~4 shows the SED of the companion in the rest frame, with the 800 Myr
model superposed on the photometry points.   Even though this model was
chosen solely on the basis of its fit to the spectroscopy at 
$\lambda_{o} < 4500$ \AA , it is in remarkable agreement with the photometry 
of the object.  The fact that the two IR points are well fit by the model 
indicates that there is little or no dust.   

We experimented with adding older stellar components, allowing both ages and 
relative contributions to vary, but we were not able to obtain any better 
fit to both the spectrum and the SED.   Therefore, at this level,
we find no evidence for a significant older stellar component in the 
companion galaxy.

\section{Discussion}

With our confirmation that the object south-southwest of UN\,J1025$-$0040
is a companion galaxy and the evidence
that the object is physically related to the quasar,   this system joins
other examples for which it is plausible that both the quasar activity 
and the starburst may have been triggered by an interaction.
From the $I$-band magnitude for the companion given in Table 1, we
estimate an absolute magnitude $M_B=-18.3$ ($H_0=75$, $q_0=0$), so
the companion is similar to the Large Magellanic Cloud in luminosity, but
more compact ($\leq$ 1.9 kpc).

We find very different ages for post-starburst populations in the host 
galaxy (400 Myr) and the companion
galaxy (800 Myr).   These ages, of course, are somewhat uncertain.  Because 
of contamination from the quasar,
and the fact that the SED of the quasar itself remains unknown, it is difficult
to model the stellar population with high accuracy.   The spectrum of the 
companion, on the other hand, is too noisy for detailed modeling.   However, 
the contrast between the strong Balmer lines and the Balmer discontinuity in
the host, and the 4000 \AA\ break in the companion, show clearly 
that the post-starburst ages of these objects cannot be the same and,
in fact, must differ by a few $\times 10^{8}$ years.
High-spatial-resolution spectroscopy could separate the spectrum of the quasar
nucleus from that of the starburst and allow a more precise determination of
the age of the starburst.

It is possible that the starburst in the companion
may have been triggered during a previous passage, while the corresponding
starburst in the host galaxy, if present, may be masked by the more recent 
starburst.  This suggestion is appealing because the orbital period of the
pair should be of the order of a few $\times 10^8$ years, while it is
difficult to imagine internal galactic processes having similar time scales 
which could cause massive starbursts; and it is equally difficult to imagine
that these two starbursts are completely unrelated.
An orbital origin for the starbursts also fits well
with the episodic star formation at times of close passage seen in the
merger models of Mihos \& Hernquist (1996\markcite{mih96}).  
At this stage, however, attributing the age difference to the
orbital period of UN\,J1025$-$0040 and its companion can only be speculation.  
It could be, for example, that the recent starburst in the host galaxy is
due to the merger of a second companion and is unrelated to the one we can
see; or the stellar populations in the host galaxy and the companion might
actually be the same age, if the initial mass function in the companion
had a sharp cutoff at about 2 solar masses.

Nevertheless, UN\,J1025$-$0040 is a key object for our attempts to understand
the starburst---AGN connection. It is the only object for which we know
that there have been recent major starbursts in both a QSO host galaxy and 
its companion,
and for which we can compare the starburst ages of each. It is the clearest
example of a ``transition'' object between a starburst galaxy and a classical
QSO. The most important task now is to determine what {\it kind} of transition
is relevant to such objects: is it an evolutionary transition, in the sense
that the objects progress from starburst to QSO (\eg\ Sanders \etal\ 
\markcite{san88}1988); or is it simply an example
of the range of properties such objects can have due to the range of
physical conditions under which they are produced? The answer to this
question depends on the relative times at which the starburst and the
QSO activity are initiated and their relative luminosities as they age
(Stockton \markcite{sto99}1999).  The largest uncertainty remains our lack of
knowledge of the luminosity history and lifetimes of QSOs. It is in this
area that close studies of objects like UN\,J1025$-$0040 may be helpful.
If we can assume (or, better, demonstrate) that the QSO activity is triggered
roughly simultaneously with the peak of the starburst, then we can conclude
that the QSO lifetime (either continuous or episodic) can be as long as
$\sim4\times10^8$ years.  If the QSO luminosity were roughly constant over this
period, then at some earlier time, the starburst would have swamped the
QSO flux, quite aside from any effect of dust. For example, when the starburst 
was $\sim50$ Myr old, it would have been $\sim2.5$ times more luminous over
most of the optical, rising rapidly to $\sim6.5$ times more luminous
between 4000 \AA\ and 3000 \AA\ in the rest frame. It would have dominated the
QSO emission at all wavelengths from at least the near-IR to close to the
Lyman limit, and the QSO would have been detectable in the optical spectrum
only from small peaks at the positions of the strongest emission lines.

This scenario would tend to support the evolutionary view of transition
objects like UN\,J1025$-$0040 and 3C\,48 (Canalizo \& Stockton 
\markcite{can00a}2000a), but it depends on assumptions about the
initiation and timescale of QSO activity.  These assumptions can only be
checked by examining post-starburst ages and nuclear activity in a
sample of objects, spanning a range of ages.

\acknowledgments

This research was partially supported by NSF under grant AST95-29078.
Data presented herein were obtained at the W. M. Keck Observatory, which is
operated as a scientific partnership among the California Institute of
Technology, the University of California, and the National Aeronautics and
Space Administration.  The Observatory was made possible by the
financial support of the W. M. Keck Foundation. 
This research has been supported in part by NSF under grant AST95-29078,
and in part
performed under the auspices of the U.S. Department of Energy
by Lawrence Livermore National Laboratory under Contract W-7405-ENG-48.

\newpage

\begin{figure}
\plotone{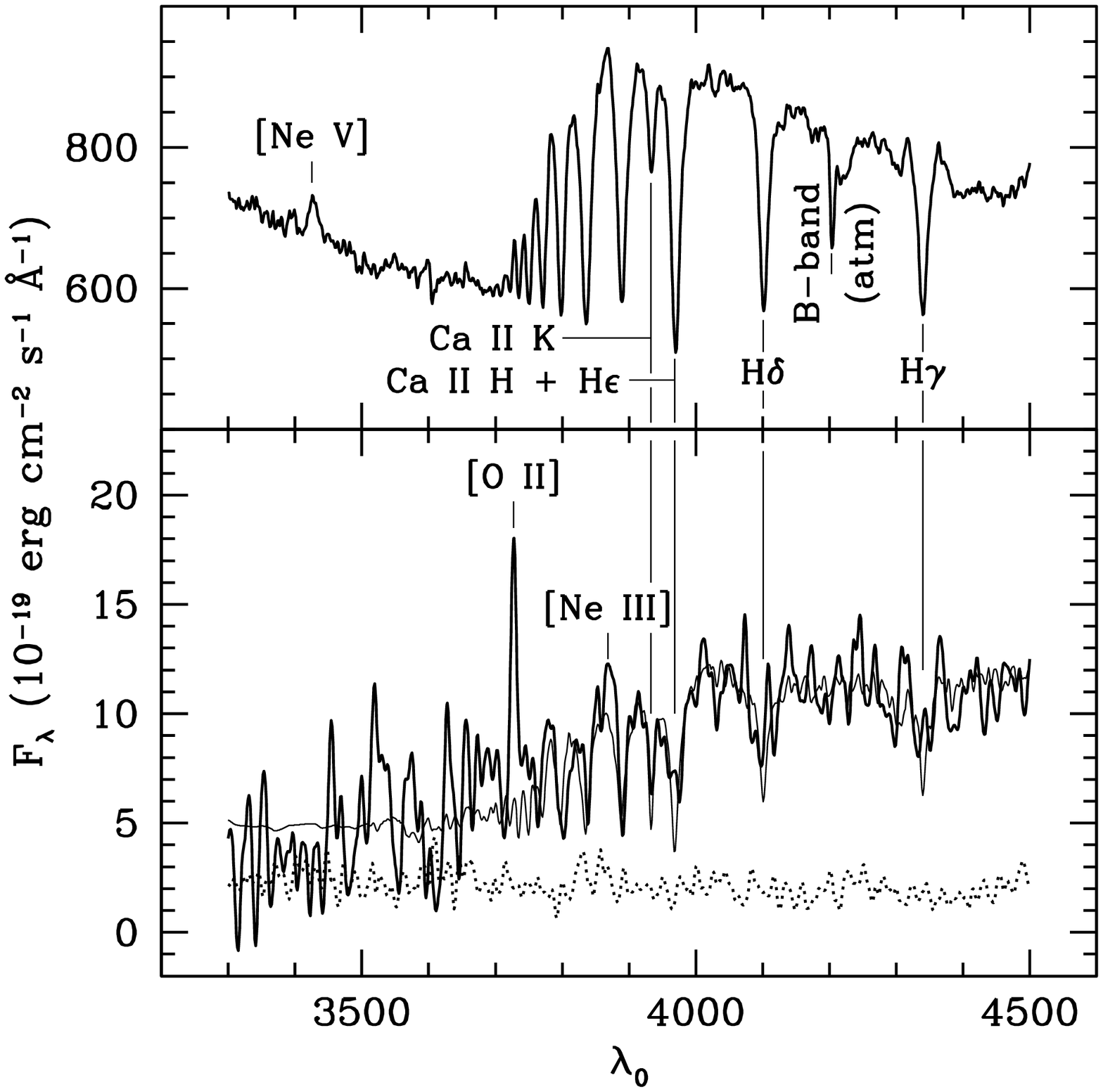}
\caption{The upper panel shows the combined spectrum of the quasar and its
post-starburst host galaxy.  The lower panel shows the spectrum of the 
companion (heavy solid line), with a Bruzual \& Charlot (1996) 800 Myr 
population model superposed (light solid line).  The dotted trace at the 
bottom shows the 1$\sigma$ random uncertainty for the observed spectrum, 
taking into account the $\sigma=3$ pixel Gaussian smoothing that has been 
applied.}
\end{figure}

\newpage
\begin{figure}
\caption{UN\,J1025$-$0040 and companion galaxy.  Sum of an $R$- and an 
$I$-band image, with total exposure of 13500s.  North is up and East is
to the left.}

\caption{Two-dimensional spectral image of UN\,J1025$-$0040 (top) and 
companion galaxy (bottom) showing the [O\,II] $\lambda3727$ emission line.
The image has been smoothed with a $\sigma=1$ pixel Gaussian filter.}
\end{figure}

\newpage
\begin{figure}
\plotone{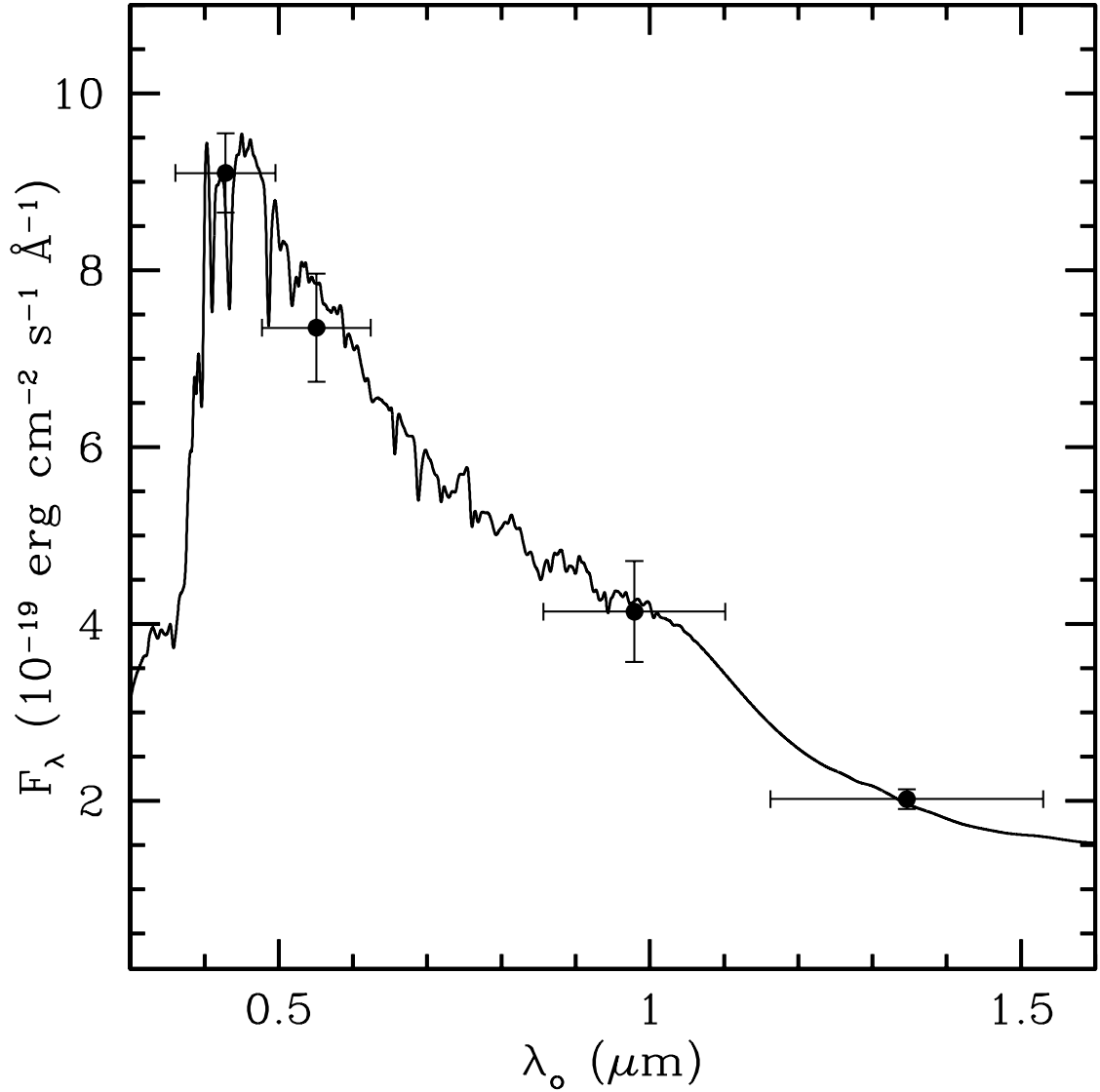}
\epsscale{1.0}
\caption{Photometry of the UN\,J1025$-$0040 companion galaxy in rest frame.
The error bars in the horizontal direction represent the de-redshifted
bandpasses, while those in the vertical direction are 1$\sigma$ random
uncertainty in the flux.   The trace is the same model as that shown 
in Fig. 1.}
\end{figure}

\newpage
\begin{deluxetable}{ccc}
\tablewidth{0pt}
\tablecaption{Photometry of UN J1025$-$0040 and its Companion}
\tablehead{\colhead{Band} & \colhead{UN J1025$-$0040} & \colhead{Companion}}
\startdata
$R$      & $18.82\pm0.02$  & $23.22\pm0.06$ \nl
$I$      & $18.40\pm0.01$  & $22.64\pm0.09$ \nl
$H$      & $16.81\pm0.01$  & $21.04\pm0.14$ \nl
$K_{s}$  & $16.67\pm0.01$  & $20.70\pm0.06$ \nl
\enddata
\end{deluxetable}

\end{document}